\documentclass[twocolumn,tighten,twocolappendix]{aastex701}

\usepackage{CJKutf8}
\graphicspath{{./}{figures/}}

\begin{document}

\title{Evidence for a Delayed Progenitor Population for CHIME non-repeating Fast Radio Bursts from a Self-Consistent Forward and Backward Inference Framework}

\begin{CJK*}{UTF8}{gbsn}

  \author[orcid=0000-0002-9110-4336]{Zi-Liang Zhang (张子良)}
  \affiliation{The Hong Kong Institute for Astronomy and Astrophysics (HKIAA), The University of Hong Kong, Pokfulam Road, Hong Kong, People's Republic of China}
  \affiliation{Department of Physics, The University of Hong Kong, Pokfulam Road, Hong Kong, People's Republic of China}
  \email[show]{zi-liang.zhang@connect.hku.hk}

  \author[orcid=0000-0002-9725-2524]{Bing Zhang (张冰)}
  \affiliation{The Hong Kong Institute for Astronomy and Astrophysics (HKIAA), The University of Hong Kong, Pokfulam Road, Hong Kong, People's Republic of China}
  \affiliation{Department of Physics, The University of Hong Kong, Pokfulam Road, Hong Kong, People's Republic of China}
  \email[show]{bzhang1@hku.hk}

  \begin{abstract}
    Fast radio bursts (FRBs) are luminous extragalactic radio transients whose physical origins remain uncertain. Using over one thousand non-repeating events from CHIME/FRB Catalog 2, we infer the intrinsic FRB demographics with a self-consistent framework that combines backward non-parametric inference and forward population synthesis while accounting for a probabilistic dispersion measure--redshift relation, baseband-to-catalog fluence corrections, and an injection-calibrated multidimensional selection function.  We first apply a backward non-parametric method, the weighted Lynden--Bell $C^{-}$ estimator, to recover the intrinsic redshift and energy distributions without assuming any population model. Independently, we perform forward Monte Carlo population synthesis in observable dispersion measure--fluence space, treating candidate intrinsic redshift and energy distributions as population hypotheses and comparing the resulting selected synthetic catalogs with observations. We find that the intrinsic redshift distribution peaks at $z\sim1$, below the cosmic star formation history (SFH) peak at $z\sim1.7$, indicating strong tension with a pure SFH-tracking scenario, suggesting that at least some FRBs are delayed with respect to SFH. The intrinsic energy distribution is consistent with a power law with index $\alpha\approx1.9$ and steepens at higher energies. We find no significant evidence for an intrinsic redshift--energy correlation.

  \end{abstract}

  \keywords{Radio bursts (1339); Radio transient sources (2008) }

  \section{Introduction}
  Fast radio bursts (FRBs) are short and intense radio transients with unknown progenitors \citep{Lorimer2007,Thornton2013,Petroff2022,Zhang2023}.
  The association of FRB 20200428A and an X-ray burst with the Galactic magnetar SGR 1935+2154 \citep[e.g.,][]{Bochenek2020,CHIME2020,HXMT2021}, which is associated with the supernova remnant G57.02+0.8 \citep{Gaensler2014,Kothes2018,Zhou2020}, points to a young environment. By contrast, FRB 20200120E, located in a globular cluster in M81, points to an old environment \citep{Kirsten2022Nature}.

  Population studies can constrain the intrinsic FRB energy function and redshift distribution, with these constraints improving as the available sample grows. \citep[e.g.,][]{Yu2014,Caleb2016,Lu2016,Cao2018a,Luo2018,Lu2019,Luo2020,Zhang2021,Zhang2022,Hashimoto2022,James2022a, Shin2023,Wang2024,Gupta2025}. Well-localized events are particularly valuable for understanding the physical nature of FRBs, but the sample size of such precisely localized sources remains small \citep[e.g.,][]{Chatterjee2017Natur,Ravi2019,Bannister2019, Marcote2020, Niu2022, Driessen2024}. Also, FRB sources were detected using a diverse array of telescopes with different observing frequencies, sensitivities, and trigger criteria. The resulting combined sample is therefore deeply heterogeneous.

  To overcome the limitations of both sample size and instrumental heterogeneity, a broader, uniformly selected sample of unlocalized FRBs provides a complementary population probe. Specifically, the intrinsic redshift distribution can place stringent constraints on FRB progenitors. If the distribution tracks the cosmic star-formation history (SFH), it implies a young progenitor population, such as magnetars formed from core-collapse supernovae. Conversely, a distribution delayed relative to the SFH implies an older population, such as magnetars formed from binary white dwarf mergers, neutron star mergers or accretion-induced white dwarf collapses. The energy function of non-repeating FRBs characterizes the distribution of burst energies across the population and therefore provides an additional demographic constraint.
  
  The Canadian Hydrogen Intensity Mapping Experiment (CHIME) \citep{CHIME2018} provides exactly this type of homogeneous dataset. The number of detected FRBs increased by an order of magnitude in the first CHIME/FRB catalog \citep{CHIME2021} relative to earlier samples. However, despite the uniform instrumental origin of these bursts, the redshift and energy distributions inferred from this catalog using different statistical methods have not converged on a common result. Through forward Monte Carlo simulations, \citet{Zhang2022} found that the FRB population is delayed with respect to the SFH. This conclusion was supported by several follow-up studies using various methods, including the backward $V_{\rm max}$ method  \citep{Hashimoto2022}, forward population modeling method \citep{Qiang2022,zzl2023}, Bayesian population analyses \citep{LinZou2024,LinLiZou2024}, the backward Lynden--Bell estimator without a fuzzy selection function \citep{Chen2024, Zhang2025, Champati2025}, and forward fitting of the baseband catalog \citep{Gupta2025}. In contrast, a population analysis based on a weighting formalism with a more careful treatment of the instrumental selection effect \citep{Shin2023} and a detailed forward population synthesis \citep{Wang2024} concluded that the data are still consistent with FRBs tracking the SFH. Although delayed models remain compatible with these analyses, they are not required by the data.

  Backward non-parametric methods provide a complementary route to FRB demographic inference because they recover the intrinsic redshift and energy distributions without imposing explicit parametric population models. In the Lynden--Bell $C^{-}$ framework, they also provide a basis for testing whether redshift and energy can be treated as statistically independent population variables\footnote{Here redshift--energy independence refers to a correlation between redshift and energy in the intrinsic population, not to the deterministic relation among fluence, luminosity distance, and isotropic-equivalent energy.} \citep{Lynden-Bell1971, Efron1992}. However, previous backward analyses generally did not include both the fuzzy CHIME/FRB selection function and the probabilistic nature of the $\mathrm{DM}_{\rm ext}$--$z$ relation \citep[e.g.,][]{Chen2024, Zhang2025, Champati2025}. As emphasized by \citet{Bryant2021}, replacing a gradual selection function with a sharp cutoff can bias the recovered intrinsic distributions and can induce apparent redshift--energy dependence. This issue is particularly relevant for CHIME/FRB, whose detection efficiency strongly depends on multiple observables rather than on a single fluence threshold \citep{CHIME2021, Merryfield2023, McGregor2026}. In addition, the catalog-reported fluence is generally a lower limit to the true fluence, which propagates directly into the inferred burst energy \citep{CHIME2021, Andersen2023, Merryfield2023, Amiri2024}.

  CHIME/FRB Catalog 2 \citep{ChimeCat2} contains more than 5000 events and provides a substantially larger sample for FRB population studies.
  In this work, we focus on non-repeating events and combine two complementary approaches that constrain the population from different directions, while explicitly propagating uncertainties associated with the ${\rm DM}_{\rm ext}$--$z$ relation and the catalog-reported fluences.
  We first convert the catalog fluences to baseband-like fluences.
  In the backward analysis, we apply the weighted Lynden--Bell $C^{-}$ method to infer intrinsic population distributions, including the peak of the redshift distribution, the slope of the energy distribution, and its steepening at high energies, and use Monte Carlo simulations to test redshift--energy independence.
  In the forward analysis, we perform Monte Carlo population synthesis as an independent population constraint in observable space: mock events generated from candidate intrinsic redshift and energy distributions are passed through the survey selection function and the ${\rm DM}_{\rm ext}(z)$ probability distribution (see Appendix~\ref{sec:appendix:redshift-energy} for the definition of $\mathrm{DM}_{\rm ext}$), and the resulting synthetic $\mathrm{DM}_{\rm ext}$ and fluence distributions are compared directly with the observed distributions.

\begin{figure*}[!ht]
    \centering
    \includegraphics[width=\textwidth]{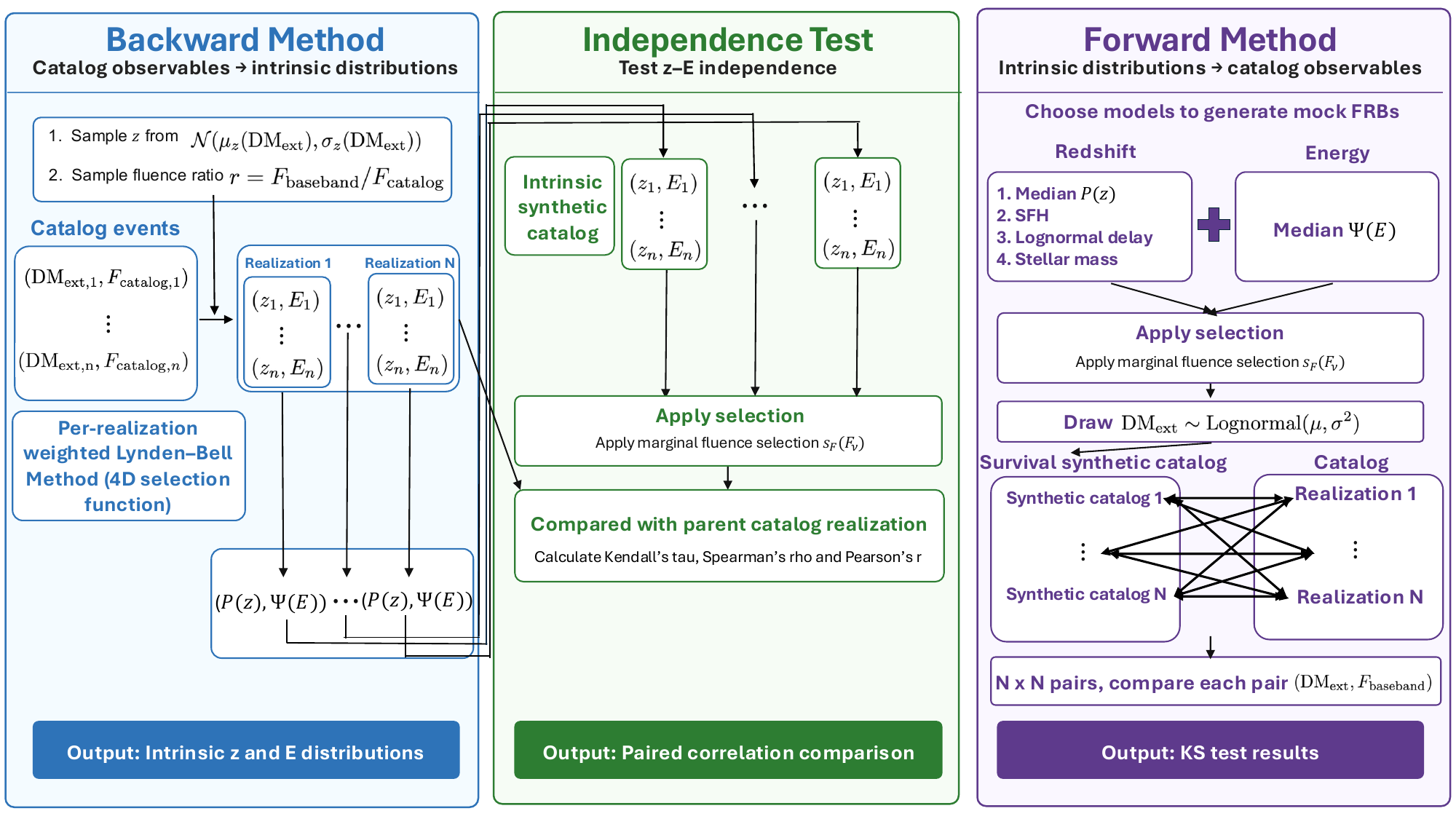} 
    \caption{Flow chart summarizing the self-consistent backward-inference and forward-synthesis framework used in this work.}
    \label{fig:flow_chart}
  \end{figure*}

  Figure~\ref{fig:flow_chart} summarizes the analysis workflow.
  Section~\ref{sec:method} describes the statistical frameworks for the weighted Lynden--Bell $C^{-}$ inference and the forward population synthesis.
  Section~\ref{sec:results} presents the inferred redshift and energy distributions, the redshift--energy independence tests, and the forward-modeling constraints.
  Section~\ref{sec:summary} summarizes the main conclusions.

  \section{Method}\label{sec:method}
  \subsection{Sample selection} \label{Sec:sample_selection}

  We construct a clean subsample from CHIME/FRB Catalog 2 \citep{ChimeCat2} by excluding
  \begin{enumerate}
    \item events detected before 2018-09-04 (commissioning phase), as well as events flagged by \textit{excluded\_flag}, \textit{citizen\_science\_flag}, or \textit{sidelobe\_flag};
    \item repeating bursts with a non-empty \textit{repeater\_name} field;
    \item highly scattered events ($\tau > 10~{\rm ms}$ at 600~MHz), which are strongly affected by CHIME selection bias;
    \item events with \textit{bonsai\_snr} $< 12$, to reduce incompleteness from human inspection near the low-S/N threshold;
    \item events with $\textit{dm\_fitb} > 2000\,{\rm pc\,cm^{-3}}$, where the selection function is not well constrained (see Figure 9 of \citet{Merryfield2023});
    \item events with absolute meridian angle larger than $1^{\circ}$, where the pattern of baseband-to-catalog fluence ratio is different from events within $1^{\circ}$.

  \end{enumerate}

  \begin{figure}[!htbp]
    \centering
    \includegraphics[width=\linewidth,clip,trim=0 0 0 4]{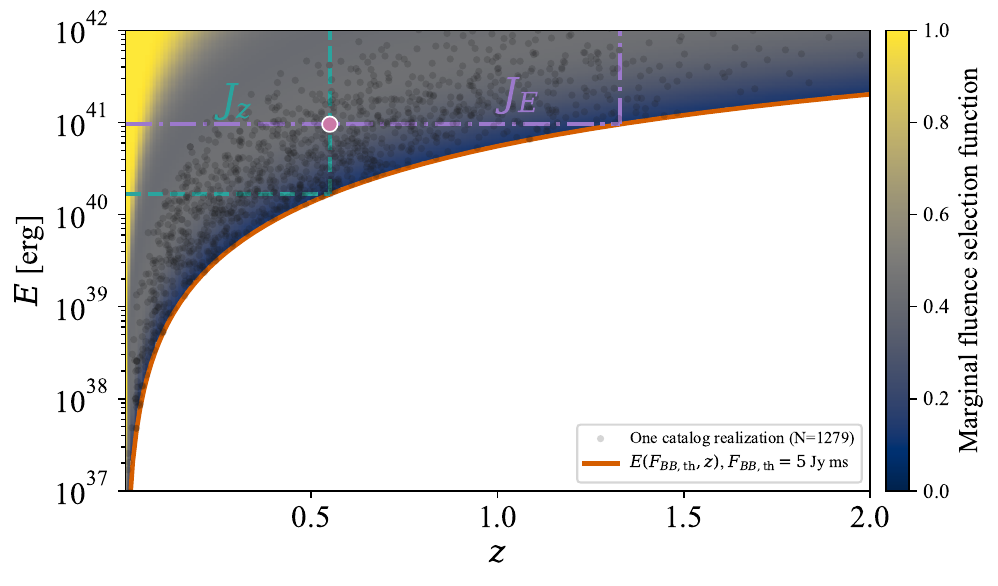}
    \caption{
      $z$--$E$ distribution for the selected CHIME/FRB Catalog 2 sample.
      The black points show the catalog events, with the colorbar indicating the marginal fluence-only projection of the selection function.
      The red solid line represents the fluence-limited selection boundary at $F_{\nu, \rm th}=5~\mathrm{Jy\,ms}$, which is used to define the comparable sets in the weighted Lynden--Bell $C^{-}$ method.
      The pink point $(z_i, E_i)$ is used to illustrate the definition of the comparable sets $J_{z_i}$ and $J_{E_i}$, which are defined by the green and purple dashed lines, respectively.
    }
    \label{fig:z_e_scatter}
  \end{figure}

  Criteria 1 through 4 represent standard quality and uniformity cuts commonly adopted in the literature \citep[e.g.,][]{CHIME2021, Hashimoto2022, Shin2023, ChimeCat2}. Criterion 5 is explicitly introduced in this work to ensure the sample is strictly confined to the parameter space where the survey selection function is well defined and robustly calibrated against injection surveys. Criterion 6 ensures that the baseband-to-catalog fluence-ratio distribution remains stable; further details are provided in the next subsection. After applying these selection criteria, a sample of 1798 non-repeating FRBs is obtained for analysis.

  \subsection{Baseband fluence and selection function} \label{sec:method:fluence-selection}
  The catalog-reported fluence generally provides only a lower limit to the burst physical fluence \citep{Andersen2023, Amiri2024, ChimeCat2}.
  Correcting the catalog fluence is required because the fluence enters directly into the energy estimate and the injection-based selection function \citep{Merryfield2023, McGregor2026} is calibrated in terms of the injected physical fluence.
  Each catalog fluence is therefore converted to a baseband-like fluence by assigning a baseband-to-catalog fluence-ratio draw from the empirical distribution of \citet{Amiri2024}\footnote{The baseband-to-catalog fluence ratio is modeled using its one-dimensional marginal distribution, without conditioning on any other burst properties. No clear trend with either the baseband-measured burst duration or scattering time is apparent within the analyzed sample.}.
  The ratio spans a broad range, from values below unity to approximately $10^3$.
  Because the empirical ratio distribution remains stable for absolute meridian angle below $1^{\circ}$ but changes systematically at larger meridian angles (see Figure 4 of \citet{Amiri2024}), fluence-ratio draws are restricted to the absolute meridian angle $<1^{\circ}$ subsample.

  The injection-based selection function is evaluated in four dimensional space \citep{McGregor2026}, $S_4(F_{\nu}, {\rm DM}_{\rm tot}, W, \tau_{600})$,
  where $F_{\nu}$ is the injection fluence, ${\rm DM}_{\rm tot}$ is the total dispersion measure, $W$ is the pulse width, and $\tau_{600}$ is the scattering time.
  We also use the marginal fluence-only projection, denoted as $S_F(F_{\nu})$, for the $z$--$E$ visualization and for the synthetic-catalog filtering in the independence and forward tests.
  To remain within the fluence range where the pipeline completeness is better constrained, only baseband-like fluences in the range $5 \le F_{\nu} \le 200~{\rm Jy~ms}$ are retained \citep{Merryfield2023,McGregor2026}. 
  After applying this draw-level cut, a typical catalog realization contains approximately 1300 events.

  \subsection{Backward Weighted Lynden-Bell \texorpdfstring{$C^{-}$}{C-} method} \label{sec:method:backward}
  Two complications affect the analysis of CHIME/FRB events: the catalog-reported fluence is generally a lower limit, and redshifts inferred from DM are probabilistic. Both effects introduce uncertainty into the inferred redshifts and energies.
  To fully account for these uncertainties, a Monte Carlo approach is adopted to infer intrinsic distributions.
  Throughout this work, a draw denotes a single sampled latent quantity, such as a redshift or fluence-ratio value for one burst. A catalog realization denotes one Monte Carlo instantiation of the latent redshifts, fluence ratios, and energies for all selected CHIME/FRB Catalog 2 events. By contrast, a synthetic catalog denotes a forward-simulated FRB population drawn from an assumed intrinsic distribution and filtered through the survey selection function.
  We propagate these uncertainties using the following Monte Carlo procedure:
\begin{enumerate}
  \item For each FRB, a latent redshift draw is sampled from its associated Gaussian distribution $z \sim \mathcal{N}(\mu_z, \sigma_z)$, and a baseband-to-catalog fluence-ratio draw is sampled to compute the corresponding energy draw $E$. The details of redshift and energy estimation are described in Appendix~\ref{sec:appendix:redshift-energy}   \footnote{
    We use Planck 2018 cosmology parameters \citep{Planck2018} in this work for comparison with previous studies, but since FRBs are local Universe phenomena and are mostly at $z\lesssim2$, SH0ES cosmology parameters should be chosen.
  Therefore, we test using the SH0ES cosmology parameters from \citet{Riess2022} and find no material difference for the result. }.
  \item We repeat this sampling $1000$ times per event, generating an ensemble of $1000$ catalog realizations.
  \item For each of these $1000$ catalog realizations, we apply the weighted Lynden--Bell method to derive an independent cumulative distribution for both redshift and energy.
\end{enumerate}

  We use the weighted Lynden--Bell $C^{-}$ method to infer the intrinsic redshift and energy distributions of FRBs \citep{Lynden-Bell1971}. This approach extends the classical non-parametric Lynden--Bell estimator by incorporating the fuzzy four dimensional selection function.

  For redshift, the comparable set \citep{Lynden-Bell1971,Efron1992} is defined as
  \begin{equation}
    J_{z_i} = \{j: z_j \le z_i,\ E_j \ge E(F_{\nu, \rm th}, z_i)\},
  \end{equation}
  as illustrated by the green dashed boundary in Figure \ref{fig:z_e_scatter}.

  For energy, the comparable set is defined as
  \begin{equation}
    J_{E_i} = \{j: E_j \ge E_i,\ z_j \le z(F_{\nu, \rm th}, E_i)\},
  \end{equation}
  as illustrated by the purple dashed boundary in Figure \ref{fig:z_e_scatter}.

  Here, $E(F_{\nu, \rm th}, z_i)$ and $z(F_{\nu, \rm th}, E_i)$ describe the same fluence-limited boundary at $F_{\nu, \rm th}=5~\mathrm{Jy\,ms}$. 
  Notice that $F_{\nu, \rm th}$ is not treated as a hard cutoff selection function like in previous non-parametric methods, but as a lower fluence boundary used to avoid CHIME pipeline noise misclassification \citep{Merryfield2023, McGregor2026}.
  $E(F_{\nu, \rm th}, z_i)$ is the minimum energy at redshift $z_i$ corresponding to this fluence boundary, and $z(F_{\nu, \rm th}, E_i)$ is the maximum redshift for a burst of energy $E_i$ at this fluence boundary. Therefore, $J_{z_i}$ compares events bright enough to be observed at $z_i$, whereas $J_{E_i}$ compares events close enough to be observed at fixed $E_i$.
  $E(F_{\nu, \rm th}, z_i)$ and $z(F_{\nu, \rm th}, E_i)$ extend the original selection boundary $\{j: E_j \ge E_i,\ z_j \le z_i\}$, thereby incorporating more events into the comparable sets and improving the precision of the inferred distributions.

  Let $F_{\nu}(z,E)$ be the baseband-like fluence corresponding to a burst with redshift $z$ and energy $E$.
  Equation (8) of \citet{Lynden-Bell1971} is written in discrete weighted form as
  \begin{equation}
    C^{-}_{z}(z_{i}) =
    \sum_{j \in J_{{z_i}}}
    \frac{
      S_4\!\left[F_{\nu}(z_i,E_j),{\rm DM}_{j\rightarrow i},W_j,\tau_{600,j}\right]
    }{
      S_4\!\left[F_{\nu}(z_j,E_j),{\rm DM}_{{\rm tot},j},W_j,\tau_{600,j}\right]
    } - 1
  \end{equation}
  where $S_4$ is the four dimensional selection function defined in Section~\ref{sec:method:fluence-selection}.
  The denominator is the selection probability of event $j$ at its observed coordinates.
  The numerator evaluates the same event after moving it to the trial redshift $z_i$ while keeping $E_j$, $W_j$, and $\tau_{600,j}$ fixed.
  The shifted total DM is defined as ${\rm DM}_{j\rightarrow i} = {\rm DM}_{{\rm tot},j} + 831(z_i-z_j)$, 
  where 831 is the slope of the median $\mathrm{DM}_{\rm ext}$--$z$ relation used in Appendix~\ref{sec:appendix:redshift-energy}.
  This shift keeps the event-specific DM residual fixed while evaluating the selection probability at the trial redshift.
  Setting \(S_4\equiv1\) reduces the weighted estimator to the original \(C^{-}\) method, which is the Kaplan--Meier estimator in survival analysis.

  The inverse denominator is the weight of each event $j$ in the comparable set, which accounts for the fact that events with lower detection probability should be weighted more heavily.
  The numerator accounts for the fact that the actual detected events at the infinitesimal redshift bin around $z_i$ are the intrinsic events at this redshift bin multiplied by the selection function at this redshift bin.
  The $-1$ is to exclude the event $i$ itself from the comparable set, which is the same trick as the original $C^{-}$ method.
  The cumulative redshift distribution is therefore

  \begin{equation}
    P(z_i) = \int_{\le z_i} p(z)\,dz = \prod_{j \le i} \left(\frac{C^{-}_{z}(z_{j}) + 1}{C^{-}_{z}(z_{j})}\right)
  \end{equation}
  where $p(z)$ is the redshift probability density function. The cumulative distribution is a product of the contributions from each $C^{-}$ of each event.

  Applying the same procedure in energy space, we obtain

  \begin{equation}
    C^{-}_{E}(E_{i}) =
    \sum_{j \in J_{{E_i}}}
    \frac{
      S_4\!\left[F_{\nu}(z_j,E_i),{\rm DM}_{{\rm tot},j},W_j,\tau_{600,j}\right]
    }{
      S_4\!\left[F_{\nu}(z_j,E_j),{\rm DM}_{{\rm tot},j},W_j,\tau_{600,j}\right]
    } - 1
  \end{equation}
  In this case the trial energy changes the fluence of event $j$, but the redshift, total DM, width, and scattering time are fixed at the observed values.

  The corresponding cumulative energy distribution is

  \begin{equation}
    \Psi(E_i) = \int_{\ge E_i} \psi(E)\,dE = \prod_{j \le i} \left(\frac{C^{-}_{E}(E_{j})}{C^{-}_{E}(E_{j}) + 1}\right)
  \end{equation}
  where $\psi(E)$ is the energy probability density function. The cumulative distribution is a product of the contributions from each $C^{-}$ of each event.

  \subsection{Simulation-based redshift--energy independence test} \label{sec:simulation test method}
  The Lynden--Bell $C^{-}$ method assumes statistical independence between redshift and energy \citep{Lynden-Bell1971}. Under this assumption, the redshift and energy distributions can be inferred as separate marginal distributions, but the joint redshift--energy distribution is not directly recovered. The same pair of marginal distributions can be projection from many joint distributions with different redshift--energy correlation structures. Testing redshift--energy independence is therefore necessary both for interpreting the physical origin of FRBs and for assessing whether the $C^{-}$ framework is self-consistent for this application.

  Although \citet{Efron1992} developed a non-parametric independence test for samples with a sharp truncation boundary, no direct analogue is available for the fuzzy selection function used here. In the weighted comparable sets, each data point carries a selection-dependent weight, so the rank structure assumed by the standard test is no longer preserved. A simulation-based independence test is therefore adopted:
  \begin{enumerate}
    \item For each catalog realization, the intrinsic redshift and energy marginal distributions are inferred with the weighted Lynden--Bell $C^{-}$ method (see Section \ref{sec:method:backward}).
    \item Intrinsic synthetic catalogs are generated by independently drawing redshift and energy values from the inferred marginal distributions.
    \item The marginal fluence selection function is applied to each synthetic catalog.
    \item The observed and synthetic samples are restricted to the same redshift and energy ranges.
    \item Pearson, Spearman, and Kendall correlation statistics are compared between the catalog realizations and the synthetic catalogs.
  \end{enumerate}

  In the third step, we apply only the marginal fluence selection function.\footnote{As a robustness check, we also repeated the backward analysis using only the marginal fluence selection function. The inferred redshift and energy distributions do not differ materially from those obtained with the four dimensional selection function. Within the present sample and model, this comparison indicates that the results are driven primarily by the adopted DM--redshift relation and fluence-dependent selection, while the additional dependence on dispersion measure, scattering time, and pulse width has a smaller effect.} The synthetic catalogs contain only the two intrinsic variables, redshift and energy. Applying the full four dimensional selection function would require assigning each synthetic burst a total dispersion measure, width, and scattering time, as well as specifying their distributions and correlations with redshift and energy. Introducing these properties would require additional assumptions about the underlying FRB population.
  
  Under the independence hypothesis, the distributions of correlation coefficients obtained from the catalog realizations and the selected synthetic catalogs should be statistically consistent.

  \subsection{Forward population synthesis method} \label{method:forward}

  Monte Carlo population synthesis provides an independent forward constraint on the intrinsic FRB population by testing candidate redshift and energy distributions directly against the observed extragalactic dispersion measure and fluence distributions. The main procedure follows standard FRB population-synthesis approaches \citep{Zhang2021, Zhang2022, Qiang2022, zzl2023} and is implemented as follows:
  \begin{enumerate}
    \item A synthetic catalog is generated by drawing redshift and energy pairs, $(z,E)$, from the intrinsic probability density functions.
    \item For each synthetic event, the baseband-like fluence $F_{\nu}$ is calculated from its drawn energy and redshift. For the same reason discussed in Section~\ref{sec:simulation test method}, we avoid introducing additional population assumptions and filter the synthetic events using the marginal fluence selection function.
    \item For each selected synthetic event, an extragalactic dispersion measure is drawn from the extrapolated $\mathrm{DM}_{\rm ext}$--redshift relation derived by \citet{Zhuge2026}. We write the median relation and its empirical upper $1\sigma$ linear offset as $\mu_{\rm DM}(z)=831z+111$ and $\sigma_{\rm DM}(z)=59z+155$, in units of ${\rm pc\,cm^{-3}}$. To account for uncertainty in this relation, $\mathrm{DM}_{\rm ext}$ at fixed redshift is drawn from a log-normal distribution, which preserves the non-negative physical boundary and allows a long high-$\mathrm{DM}_{\rm ext}$ tail from line-of-sight density fluctuations \citep{McQuinn2014, Macquart2020}. The draw is implemented with \texttt{numpy.random.lognormal}.
    \item For each catalog realization, baseband-like fluences are recomputed by drawing a fresh baseband-to-catalog fluence ratio for each observed FRB from the same empirical distribution used in Section~\ref{sec:method:backward}. These fluence-ratio draws are independent of those used in the backward inference.
    \item Each catalog realization is compared with each selected synthetic catalog using two-sample Kolmogorov--Smirnov (KS) tests applied separately to the $\mathrm{DM}_{\rm ext}$ and baseband-like fluence distributions.
  \end{enumerate}

  \section{Results}\label{sec:results}

  \subsection{Marginal redshift and energy distribution} \label{Sec_redshift}

  \begin{figure}[!htbp]
    \centering
    \includegraphics[width=\linewidth,clip,trim=0 0 0 4]{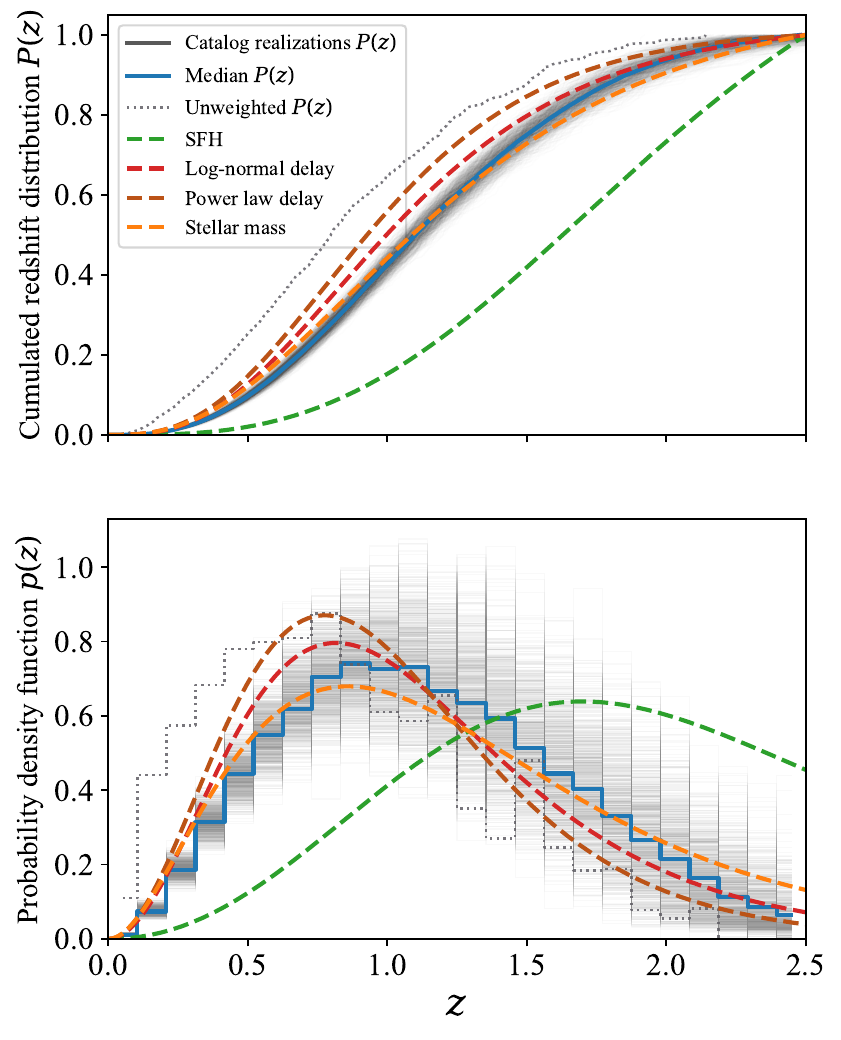}
    \caption{Redshift distribution functions.
      Upper panel: cumulative distribution functions.
      Gray solid curves show the cumulative distributions derived from the 1000 catalog realizations, and the blue solid curve shows their median.
      The green dashed curve denotes the SFH model; the red dashed curve denotes a delayed star-formation model with a log-normal delay time distribution with $\tau_{\rm LN}=5~{\rm Gyr}$ and $\sigma_{\rm LN}=1$; the brown dashed curve denotes a delayed star-formation model with a power-law delay time distribution with $\tau_c=5~{\rm Gyr}$; and the orange dashed curve denotes the stellar mass density model.
      The gray dashed curve shows the unweighted $C^{-}$ result obtained using the original catalog fluence, without the baseband-to-catalog fluence-ratio corrections.
      Lower panel: probability density functions.
      The same color scheme is used as in the upper panel.
      The gray solid histogram shows the probability density function derived from the weighted catalog realizations, while the gray dashed histogram shows the corresponding unweighted result.}
    \label{fig:Distz}
  \end{figure}

  The cumulative redshift distributions inferred from the 1000 CHIME/FRB Catalog 2 realizations are shown by the gray solid curves in the upper panel of Figure \ref{fig:Distz}.
  The corresponding probability density function is shown by the gray solid histogram in the lower panel.
  The inferred redshift distribution peaks near $z\sim1$, below the SFH peak at $z\sim1.7$.
  In both panels, the inferred distribution departs substantially from the SFH model and is closer to delayed population models.
  Details of the model redshift distributions are given in Appendix~\ref{sec:appendix:models}.
  The stellar mass density model (orange dashed curve) lies within the uncertainty region.
  The representative delayed models shown in Figure \ref{fig:Distz} include a log-normal delay time distribution with $\tau_{\rm LN}=5~{\rm Gyr}$ and $\sigma_{\rm LN}=1$, and a power-law delay time distribution with $\tau_c=5~{\rm Gyr}$.
  
  These comparisons are intended as qualitative references rather than formal best fits. Their quantitative interpretation is limited by two sources of uncertainty. First, the inferred redshift distribution remains sensitive to uncertainties in the $z(\mathrm{DM}_{\rm ext})$ relation and the empirical baseband-to-catalog fluence-ratio correction. Second, an effective delay-time distribution does not generally map uniquely to a progenitor-evolution pathway. We therefore use these phenomenological models only to determine whether the recovered FRB redshift distribution more closely resembles prompt SFH tracking or a broadly delayed evolutionary history. The adopted delay-time parameters should not be interpreted as best-fit measurements or as unique constraints on FRB progenitor evolution.

  As an additional check on the fluence-ratio correction, Appendix~\ref{baseband result} presents the marginal redshift distribution inferred from the baseband catalog, for which fluences are directly measured.
  Its agreement with the weighted Catalog 2 inference supports the use of the empirical fluence-ratio correction.

\begin{figure}[!htbp]
    \centering
    \includegraphics[width=\linewidth,clip,trim=0 0 0 4]{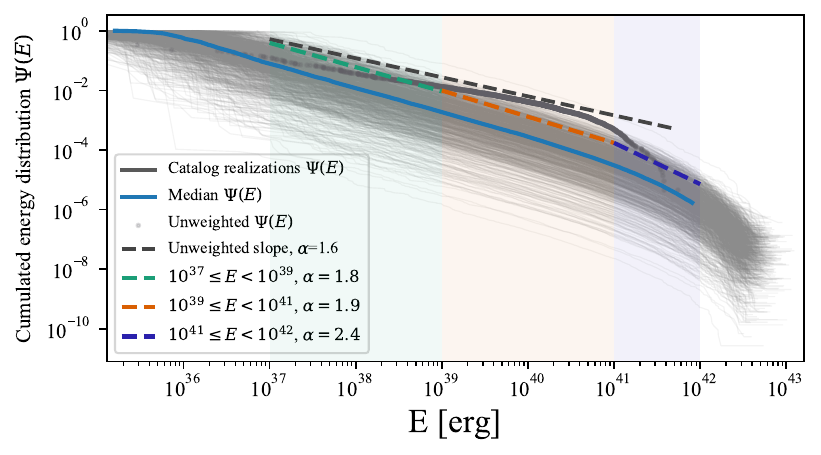}
    \caption{Cumulative energy distribution functions.
    Gray solid curves show the cumulative distributions derived from the 1000 catalog realizations, and the blue solid curve shows their median.
    $\alpha$ denotes the slope of the power-law energy probability density function, $\psi\propto E^{-\alpha}$.
    Dashed lines show piecewise power-law fits to the cumulative distribution, with different segments indicated by different colors.
    Gray points show results from the unweighted original $C^{-}$ estimator with the original catalog fluence, which yields a shallower slope.
    }

    \label{fig:cumE}
  \end{figure}

  The cumulative energy distribution is shown in Figure \ref{fig:cumE}.
  No distinct feature is found near $E\sim10^{39}~{\rm erg}$.
  Assuming a power-law probability density, $\psi\propto E^{-\alpha}$, the inferred slope is $\alpha\approx1.9$ below $10^{42}~{\rm erg}$.
  The distribution steepens, broadly consistent with previous studies \citep[e.g.,][]{Luo2020,Hashimoto2022,Shin2023}.

  Setting the selection function to unity recovers the unweighted original $C^{-}$ estimator.
  For comparison with previous unweighted results \citep[e.g.,][]{Chen2024, Zhang2025, Champati2025}, the baseband-to-catalog fluence ratio is also fixed to unity and the redshift is only the median value of ${\rm DM}_{\rm ext}$--$z$ relation without propagating its probability distribution.
  The corresponding redshift results are shown by the gray dashed curve and histogram in Figure \ref{fig:Distz}, while the corresponding energy results are shown by the gray points in Figure \ref{fig:cumE}.
  The unweighted inference yields a more delayed redshift distribution than the weighted cases and a shallower energy slope of approximately $\alpha\approx1.6$.

  \subsection{Redshift--energy independence test}

  Although the inferred energy distribution shown in Figure~\ref{fig:cumE} spans approximately $10^{35}$--$10^{43}~{\rm erg}$, the low-energy tail below $10^{37}~{\rm erg}$ is constrained by relatively few data points.
  We therefore impose $E_{\rm min}=10^{37}~{\rm erg}$ in the independence-test simulations, restricting the synthetic draws to the energy range where the inferred distribution is more densely sampled and more stable.
  This lower energy cut also reduces the computational cost of sampling a distribution that spans many orders of magnitude.
  For each independence-test realization, $3\times10^6$ intrinsic synthetic FRB events are generated.
  After the marginal fluence selection function is applied, this choice yields approximately 1300 detectable events, comparable to the final observed sample size.

\begin{figure}[!htbp]
    \centering
    \includegraphics[width=\linewidth,clip,trim=0 0 0 4]{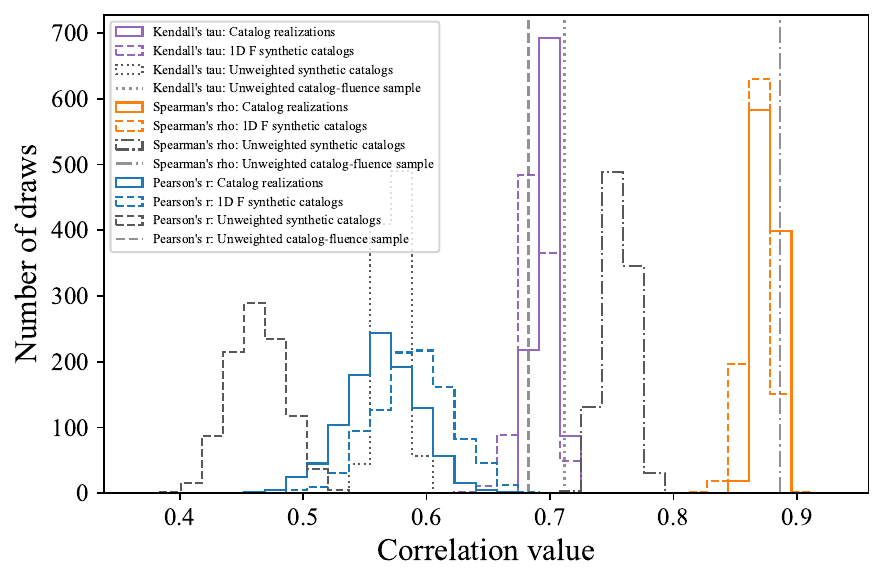}
    \caption{Correlation coefficients between redshift and energy.
    Solid histograms show the coefficient distributions from the catalog realizations.
    Dashed histograms show the corresponding distributions from 1000 synthetic catalogs generated under the redshift--energy independence assumption and passed through the marginal fluence selection function.
    Gray vertical lines show the median coefficients from the unweighted inference with the original catalog fluence, and gray dashed histograms show the corresponding unweighted synthetic-catalog distributions.}
    \label{fig:correlation}
  \end{figure}

  A total of 1000 independence-test realizations are generated under the redshift--energy independence assumption.
  Pearson, Spearman, and Kendall correlation coefficients\footnote{Computed with \texttt{scipy.stats} \citep{Virtanen2020SciPy}.} are evaluated for each realization.
  The resulting distributions are shown in Figure \ref{fig:correlation}; solid histograms denote the catalog realizations, and dashed histograms denote the synthetic catalogs generated under the independence assumption.
  For all three statistics, the catalog-realization distributions overlap substantially with the corresponding synthetic-catalog distributions.
  This agreement indicates that the data do not provide significant evidence for an intrinsic redshift--energy distribution correlation after accounting for the selection function and the baseband-to-catalog fluence correction.

  The unweighted comparison is constructed by setting the selection function to unity, fixing the baseband-to-catalog fluence ratio to unity, and assigning each burst the median redshift from the ${\rm DM}_{\rm ext}$--$z$ relation without propagating the redshift-probability distribution \citep[e.g.,][]{Chen2024, Zhang2025, Champati2025}.
  The catalog correlation coefficients for this unweighted case are shown by gray vertical lines, while the corresponding independence-test distributions from 1000 synthetic catalogs are shown by gray dashed histograms.
  The gray vertical lines are clearly displaced from the corresponding gray histograms, indicating a strong apparent redshift--energy correlation in the unweighted case.
  This apparent correlation is likely induced by selection effects \citep{Bryant2021}, because the unweighted estimator does not account for the fuzzy selection function and treats the catalog-reported fluence, which is generally a lower limit, as the intrinsic fluence.

  \subsection{Forward population synthesis }

  Monte Carlo population synthesis is used as an independent forward constraint on the population.
  $\rm DM_{\rm ext}$ and fluence distributions are directly measured observables in CHIME/FRB Catalog 2, whereas redshift and energy are inferred variables. Therefore, candidate intrinsic redshift and energy distributions can be constrained by comparing the $\rm DM_{\rm ext}$ and fluence distributions from the forward simulation with the observed $\rm DM_{\rm ext}$ and fluence distributions.
  To ensure a fair comparison with observations, each synthetic catalog realization is truncated using the same observed ranges in $\rm DM_{\rm ext}$ and baseband-like fluence (minimum and maximum values). The resulting $\rm DM_{\rm ext}$ and baseband-like fluence distributions are then compared with the observed distributions.

\begin{figure}[!htbp]
    \centering
    \includegraphics[width=\linewidth,clip,trim=0 0 0 4]{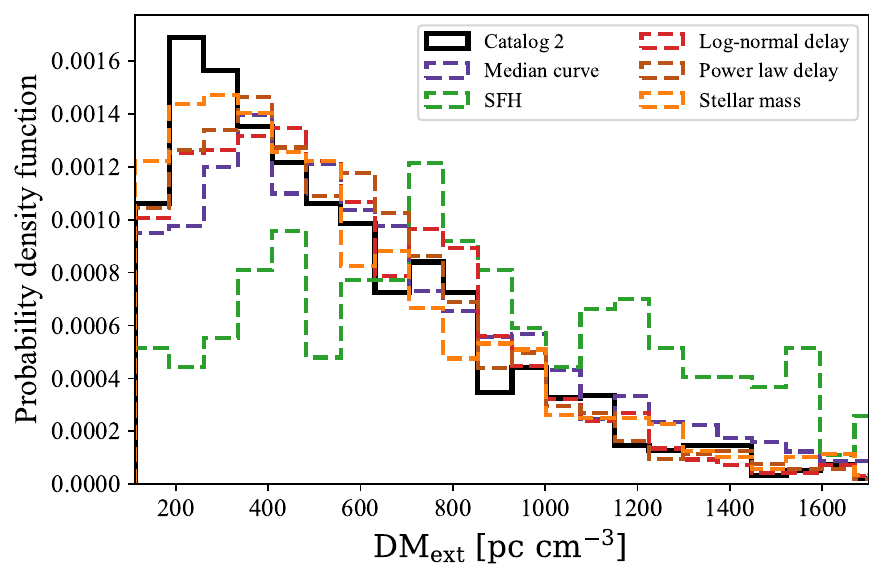}
    \caption{$\rm DM_{\rm ext}$ distribution. The same color scheme is used as in Figure \ref{fig:Distz}, except that the CHIME Catalog 2 $\rm DM_{\rm ext}$ distribution is shown by the solid black histogram.
    For each detectable synthetic event, $\rm DM_{\rm ext}$ is assigned using $\rm DM_{\rm ext}(z)$ distribution and the same $\rm DM_{\rm ext}$ selection in Section~\ref{Sec:sample_selection} is applied.
    The horizontal axis is linear, whereas \citet{Shin2023} and \citet{ChimeCat2} use a logarithmic axis. Probability densities in the two coordinates are related by \(p(x)=p(\log x)/x\). We use a linear axis because \(\mathrm{DM}_{\rm ext}\) is approximately a linear proxy for redshift.}
    \label{fig:DMdist}
  \end{figure}

  Figure \ref{fig:DMdist} compares the weighted Lynden--Bell result from CHIME/FRB Catalog 2 with representative models using the same parameter choices as in Figure \ref{fig:Distz}. The SFH model (dash green histogram) predicts a $\rm DM_{\rm ext}$ distribution that peaks at a higher value than the observed distribution, while the delayed models are more consistent with the observed distribution.

\begin{figure}[!htbp]
    \centering
    \includegraphics[width=\linewidth,clip,trim=0 0 0 4]{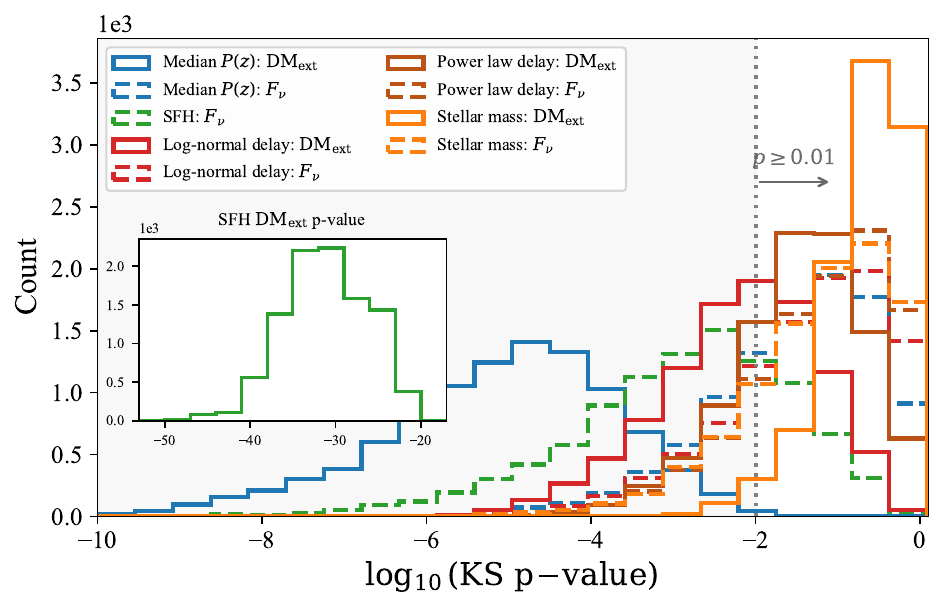}
    \caption{
      Kolmogorov–Smirnov test $p$-values for $\rm DM_{\rm ext}$ and $F_{\nu}$.
      Same color scheme is used as in Figure \ref{fig:Distz}.
      The solid histograms show the distribution of $p$-values of $\rm DM_{\rm ext}$.
      The dashed histograms show the distribution of $p$-values of baseband-like fluence.
      The zoom-in panel shows the $p$-value distribution of $\rm DM_{\rm ext}$.
    }
    \label{fig:KSp}
  \end{figure}

  Because individual synthetic-catalog realizations can fluctuate, stability is assessed from the distribution of pairwise KS-test \(p\)-values obtained by comparing the catalog realizations with the synthetic catalogs rather than from a single run.
  All models can pass the KS test for baseband-like fluence, but the delayed models are more likely to pass.
  The KS test $p$-value for the SFH model in $\rm DM_{\rm ext}$ is $\sim 10^{-30}$, indicating strong tension with the observed $\rm DM_{\rm ext}$ distribution across synthetic-catalog realizations (Figure \ref{fig:KSp}).

  Reproducing the observed \(\mathrm{DM}_{\rm ext}\) and fluence distributions therefore provides an independent consistency check on the inferred redshift and energy distributions.

  \section{Summary and Discussion}\label{sec:summary}

  In this work, we apply a backward weighted Lynden--Bell $C^{-}$ framework to the CHIME/FRB Catalog 2, explicitly using survey selection effects through an injection-based multidimensional selection function. Redshift and energy distribution independence is tested with a simulation-based approach.
  An independent forward population synthesis is also used to constrain candidate intrinsic population models against final observables. By combining this weighted non-parametric inference with forward modeling, we first provide a self-consistent demographic description of the Catalog 2 population. Crucially, our dual-methodology approach is designed to be both concise and robust: it intentionally avoids introducing excessive complexity, free parameters, or model assumptions, while strictly preserving the proper rigor required to account for complex observational selection biases.

  The main results are summarized as follows.

  \begin{enumerate}
    \item The redshift distribution peaks near $z\sim1$ and deviates from a pure SFH (SFH peak near $z\sim1.7$). The inferred shape is more consistent with delayed-population models than with SFH evolution.

    \item The energy distribution follows a power law with slope \(\alpha\approx1.9\) below \(10^{42}\,\mathrm{erg}\) and steepens at higher energies.

    \item A simulation-based independence test in the redshift--energy plane (1000 realizations) shows that the Pearson, Spearman, and Kendall coefficients of catalog realizations largely overlap with the synthetic-catalog distributions, indicating that the observations are consistent with redshift--energy independence.

    \item Taking the $\rm DM_{\rm ext}(z)$ probability distribution into consideration, forward population synthesis in observable space also indicates strong tension between the SFH model and the observed $\rm DM_{\rm ext}$ distribution (typical KS-test $p\sim10^{-30}$).
  \end{enumerate}

  Both the backward weighted Lynden--Bell inference of the intrinsic redshift distribution and the forward population synthesis of the observed $\mathrm{DM}_{\rm ext}$ distribution independently indicate that the CHIME/FRB Catalog 2 population is delayed relative to the SFH. This self-consistent result aligns with previous demographic studies utilizing various methods and earlier samples \citep[e.g.,][]{Zhang2022, Hashimoto2022, Qiang2022, zzl2023, Chen2024, LinZou2024, LinLiZou2024, Zhang2025, Gupta2025}.

  Explicitly incorporating the fuzzy survey selection function into the Lynden--Bell framework reduces several biases present in previous non-parametric studies. Compared to the weighted method, unweighted inference yields a redshift distribution that is notably more delayed. Moreover, the weighted method does not recover a significant redshift--energy correlation, highlighting the importance of accounting for selection effects in FRB demographics.

  Several methodological caveats should be noted.
  First, the baseband-to-catalog fluence ratio is sampled from an empirical distribution.
  As a consistency check, Appendix~\ref{baseband result} applies the weighted Lynden--Bell method directly to the baseband catalog, where fluences are better measured.
  The agreement between the Catalog 2 inference and the baseband-catalog inference supports the empirical fluence-ratio correction.
  Second, the backward Lynden--Bell weights use the four dimensional injection-based selection function, whereas the independence test and the forward-synthesis filtering use the marginal fluence selection function.
  This lower dimensional treatment reduces the assumptions and free parameters needed to model the dispersion-measure, pulse-width, and scattering-time distributions, but it may leave residual selection biases because the CHIME/FRB detection efficiency is coupled across fluence, dispersion measure, pulse width, and scattering time \citep{McGregor2026}.
  Appendix~\ref{baseband result} also applies the marginal fluence selection function to the baseband sample.
  The resulting redshift distribution is consistent with the inference based on the four dimensional selection function, suggesting that the marginal fluence approximation is qualitatively adequate.
  Third, the backward inference approximates the redshift uncertainty as Gaussian, which may affect the recovered intrinsic redshift distribution.
  Finally, the energy estimate in Equation~\ref{eq:energy} carries an implicit spectral assumption: it treats the relevant energy scale as a band-averaged CHIME-band quantity, appropriate if bursts are effectively narrowband or if the event rate does not vary strongly across the observing frequencies considered here.
  Appendix~\ref{appendix:varying index} examines the alternative case in which FRBs are broadband sources with a frequency-dependent spectrum.
  Unless the FRB occurrence rate is strongly frequency dependent, or the spectral index is extremely steep, this test indicates that the main redshift-distribution conclusion is not qualitatively changed.

  The inferred delay relative to the SFH is consistent with the existence of FRBs in old stellar environments, including globular clusters and early-type galaxies \citep[e.g.,][]{Bhardwaj21,Eftekhari25}, and with the finding that FRB host-galaxy properties occupy an intermediate regime between long-GRB hosts, which more closely trace star formation, and short-GRB hosts, which are more delayed \citep{LiZhang20}.
  At the same time, some FRBs, especially active repeaters, are associated with star-forming environments \citep[e.g.,][]{Tendulkar2017,Marcote2020,Niu2022,Xu2022}.
  A mixed population is therefore more plausible than a single progenitor channel.
  Because this work focuses on apparent non-repeaters, the results may indicate that this subsample is more delayed than the active-repeater population.
  In a magnetar framework, this trend would be consistent with a decline in FRB activity with source age.
  It could also indicate additional engines or triggering channels whose rates do not closely follow the SFH.
  A larger sample of precisely localized FRBs will be required to distinguish these possibilities.

\end{CJK*}

\begin{acknowledgments}

  We thank Matthew Bailes, Vicky Kaspi, and Kaitlyn Shin for helpful comments. Z.Z. is deeply grateful to Yi-Han Iris Yin for exceptionally insightful discussions on the key ideas of the Lynden--Bell method.
  Z.Z. also thanks Jiaming Zhuge for helpful discussions and for sharing code related to the DM--$z$ relation and its uncertainty distribution.
  AI-assisted tools (Codex and Gemini) were used by Z.Z. to support coding, code checking, and language polishing.
  Z.Z. reviewed and edited all AI-assisted outputs and takes full responsibility for the manuscript content, analyses, conclusions, code, and any remaining errors.
  This research has made use of NASA's Astrophysics Data System (ADS).
  The code used in this work is available from the authors upon reasonable request.

  \software{Astropy \citep{Astropy2022}, NumPy \citep{Numpy}, Matplotlib \citep{Matplotlib}, SciPy \citep{SciPy}, tqdm \citep{tqdm}}

\end{acknowledgments}

\appendix
\linenumbers

\section{Redshift and Energy Estimation} \label{sec:appendix:redshift-energy}
The observed dispersion measure $\mathrm{DM}_{\rm obs}$ of an FRB can be decomposed into Galactic and extragalactic components \citep[e.g.,][]{Thornton2013,Deng2014}:
\begin{equation}
  \mathrm{DM}_{\rm obs} = \mathrm{DM}_{\rm MW,ISM} + \mathrm{DM}_{\rm MW,halo} + \mathrm{DM}_{\rm ext},
\end{equation}
and
\begin{equation}
  \mathrm{DM}_{\rm ext} = \mathrm{DM}_{\rm diff} + \frac{\mathrm{DM}_{\rm host}}{1+z}.
\end{equation}
Here, $\mathrm{DM}_{\rm MW,ISM}$ is the Milky Way interstellar medium contribution (estimated using the YMW16 model; \citealt{YMW16}). $\mathrm{DM}_{\rm MW,halo}$ is the Milky Way halo contribution and $30 \ \mathrm{pc\,cm^{-3}}$ is adopted in this work \citep[e.g.,][]{Dolag2015}, whose uncertainty is negligible, compared to other contributions. The extragalactic term $\mathrm{DM}_{\rm ext}$ consists of the diffuse intergalactic medium and intervening halo contributions $\mathrm{DM}_{\rm diff}$, plus the redshift diluted contribution from the FRB host galaxy $\mathrm{DM}_{\rm host}$.

In the backward inference, uncertainty in the $\mathrm{DM}_{\rm ext}$--$z$ relation is propagated by treating the redshift of each FRB as a random variable. For each event, a redshift is drawn at fixed $\mathrm{DM}_{\rm ext}$ and paired with an independent draw from the empirical baseband-to-catalog fluence-ratio distribution. The corresponding isotropic-equivalent energy is then recalculated using Equation~\ref{eq:energy}. 
We evaluate the modeled distribution $p(\mathrm{DM}_{\rm ext}\mid z)$ from \citet{Zhuge2026} and extract its 16th, 50th, and 84th percentile curves. Linear fits to these curves are then inverted to obtain the median pseudo-redshift and its characteristic uncertainty at fixed $\mathrm{DM}_{\rm ext}$. The inverted median relation is
$
  \mu_z(\mathrm{DM}_{\rm ext})
  =
  \frac{\mathrm{DM}_{\rm ext}/(\mathrm{pc\,cm^{-3}})-111}{831},
$
and the difference between the inverted median and upper-$\mathrm{DM}_{\rm ext}$ percentile curves gives $\sigma_z(\mathrm{DM}_{\rm ext})$. Over the adopted range, the latter is numerically approximated by
$
  \sigma_z(\mathrm{DM}_{\rm ext})
  \simeq
  8\times10^{-5}
  \left(
    \frac{\mathrm{DM}_{\rm ext}}
         {\mathrm{pc\,cm^{-3}}}
  \right)
  +0.17.
$

At fixed redshift, the distribution of $\mathrm{DM}_{\rm ext}$ is approximately log-normal and exhibits an extended high-DM tail arising from line-of-sight density fluctuations and variations in the host-galaxy contribution \citep[e.g.,][]{McQuinn2014,Macquart2020,Zhangzj2021}. The corresponding inverse distribution of redshift at fixed $\mathrm{DM}_{\rm ext}$ is therefore asymmetric, with an extended tail toward low redshift, and does not have a simple analytic form.

For computational tractability, we approximate this inverse distribution with a Gaussian,
$
  z
  \sim
  \mathcal{N}\!\left[
    \mu_z(\mathrm{DM}_{\rm ext}),
    \sigma_z^2(\mathrm{DM}_{\rm ext})
  \right].
$
For events at low $\mathrm{DM}_{\rm ext}$, the Gaussian is truncated at $z=0$ to enforce the physical non-negative boundary. This symmetric approximation does not reproduce the full low-redshift tail of the inverse distribution and may overrepresent its high-redshift side; it is adopted as a practical means of propagating the event-level redshift uncertainty through the population inference.

In the forward population synthesis, the direction of the sampling is reversed: the redshift is drawn first from a candidate intrinsic population model, after which $\mathrm{DM}_{\rm ext}$ is drawn from the corresponding distribution, as described in Section~\ref{method:forward}.

The isotropic equivalent energy of each FRB is then calculated as
\begin{equation} \label{eq:energy}
  E = \frac{1}{1+z} 4\pi d_L^2 F_{\nu} \Delta \nu,
\end{equation}
where $d_L$ is the luminosity distance assuming Planck 2018 cosmology \citep{Planck2018}, $F_{\nu}$ is the baseband like-fluence, and $\Delta \nu = 400\,\mathrm{MHz}$ is the bandwidth of the CHIME band, since reported fluence is averaged over this bandwidth \citep{CHIME2021}. Using $\Delta\nu$ rather than the central frequency $\nu_c$ to calculate FRB energy is more appropriate for FRBs with a relatively narrow spectrum \citep{Zhang2023}. Most Catalog 2 bursts exhibit spectra that peak within the CHIME band, although some show spectra that rise or decline across the band.
The uncertainty in the derived energy is dominated by the baseband-to-catalog fluence correction and the redshift estimate. Although this calculation introduces no fitted population parameters, it depends on the adopted cosmology, bandwidth, and spectral assumptions.

\section{Redshift Distribution Models}\label{sec:appendix:models}

To physically interpret the inferred FRB redshift distribution, the results are compared against theoretical population models.

The cosmic star-formation history parameterized by \citet{Madau2014} is adopted. The pure cosmic star-formation history model assumes the FRB event rate directly tracks the star-formation rate without delay, which is characteristic of young progenitor systems (e.g., magnetars born from core-collapse supernovae). We also consider a stellar-mass-density model \citep[Equation 2 of][]{Madau2014}, which assumes the FRB rate is proportional to the total accumulated stellar mass in the Universe, representative of very old stellar populations.

Delayed models account for the evolutionary timescale between the formation of the progenitor system and the subsequent FRB event. The delayed comoving FRB event-rate density (events per unit time per unit volume) can be written as the convolution of the star-formation rate with a delay time distribution \citep[e.g.,][]{Wanderman2015,Cao2018a}:
\begin{equation}
  R_{\rm FRB}(z) \propto (1+z) \int_{0}^{t(z)-t_b} \frac{\rho_{\rm SFR}\!\left[t(z)-\tau\right]}{1+z\!\left[t(z)-\tau\right]} P(\tau)\, d\tau,
\end{equation}
where $R_{\rm FRB}(z)$ is the comoving FRB event-rate density, $\rho_{\rm SFR}$ is the cosmic star-formation rate density, $P(\tau)$ is the delay time distribution, and $\tau$ is the delay time. Equivalently, integrated over redshift, this becomes:
\begin{equation}
  R_{\rm FRB}(z) \propto (1+z) \int_{z}^{z_b} \frac{\rho_{\rm SFR}(z')}{1+z'} P\!\left[t(z)-t(z')\right] \left|\frac{dt}{dz'}\right|dz'.
\end{equation}
The maximum redshift of star formation is set to $z_b=10$. Because the cosmic star-formation rate drops rapidly at $z > 2$, the exact choice of $z_b$ does not meaningfully affect the resulting convolution.

The delay time distribution, $P(\tau)$, characterizes the probability of a FRB event occurring at a given time $\tau$ after progenitor formation. This delay is modeled using two commonly adopted functional forms \citep[e.g.,][]{Wanderman2015,Cao2018a}:

(i) A log-normal function, defined as:
\begin{equation}
  P(\tau) \propto \exp\!\left[-\frac{\left(\ln\tau-\ln\tau_{\rm LN}\right)^2}{2\sigma_{\rm LN}^2}\right].
\end{equation}

 (ii) An inverse-cutoff power-law function, defined as \citep{Cao2018a}:
\begin{equation}
  P(\tau) \propto \left(\frac{\tau}{\tau_c}\right)^{-1}\exp\!\left(-\frac{\tau_c}{\tau}\right).
\end{equation}

 For $\tau\gg\tau_c$, this function approaches $P(\tau)\propto\tau^{-1}$ \citep[e.g.,][]{Piran1992, Ruiter2009, Maoz2012}. For $\tau\ll\tau_c$, the exponential factor $\exp(-\tau_c/\tau)$ suppresses the probability of short delays, producing a smooth turnover rather than a sharp cutoff.

Finally, to convert the comoving volume density to an observable redshift distribution, the probability density function is given by \citep[e.g.,][]{Zhang2021}:
\begin{equation}
  p(z) \propto R_{\rm FRB}(z) \frac{dV_c}{dz} \frac{1}{1+z},
\end{equation}
where $dV_c/dz$ is the comoving volume element per unit redshift, and the $1/(1+z)$ factor accounts for cosmological time dilation.

\section{Baseband Catalog Consistency Check} \label{baseband result}
To test the empirical baseband-to-catalog fluence correction, the weighted Lynden--Bell $C^{-}$ method is also applied directly to the original baseband catalog \citep{Amiri2024}.
The baseband catalog has much higher time resolution than Catalog 2, which can lead to different measured widths and scattering times for the same burst.
For this reason, only the marginal fluence selection function, $S_F(F_{\nu})$, is applied to the baseband sample.
The same sample selection criteria as Section~\ref{Sec:sample_selection} are applied to ensure a consistent comparison.

\begin{figure}[!htbp]
  \centering
  \includegraphics[width=\linewidth,clip,trim=0 0 0 4]{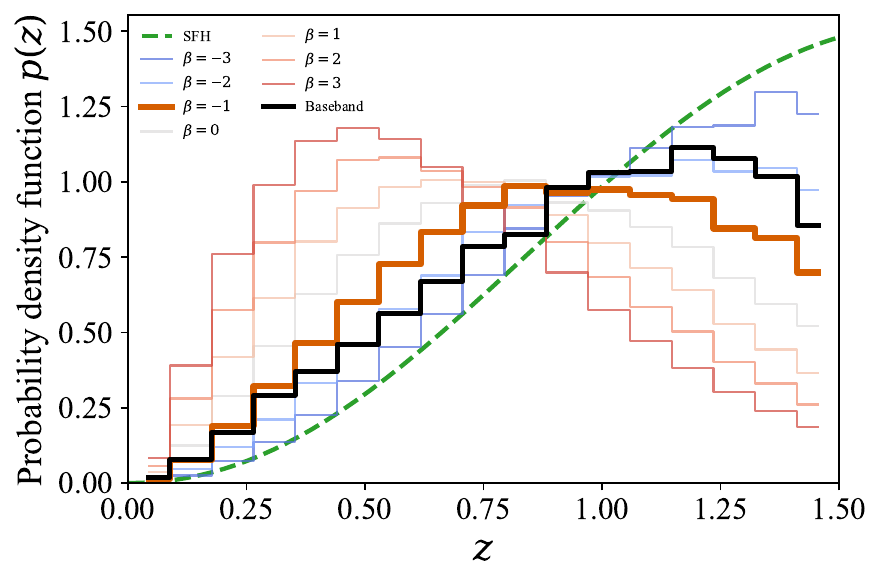}
  \caption{Redshift distribution functions from the baseband catalog and from tests with different spectral indices.
  The baseband-catalog result is shown by the black solid line. After applying the same sample selection as Section~\ref{Sec:sample_selection}, 64 baseband bursts remain. The marginal fluence selection function is applied to the baseband catalog.
  Colored solid lines show the inferred redshift distributions for different spectral indices. The $\beta=-1$ case corresponds to the main-text result.
  The SFH model is shown by the dashed green curve.
  Because of the limited baseband sample size and for a fair comparison, all distributions are truncated at $z=1.5$ and normalized by the area under the curve.
  }
  \label{fig:Distz_baseband}
\end{figure}

The inferred redshift distribution is shown in Figure~\ref{fig:Distz_baseband}.
The resulting redshift distribution is consistent with the main text inference based on the four dimensional selection function. This agreement indicates that neither the empirical baseband-to-catalog fluence correction nor the use of the marginal fluence selection function drives the inferred delay.

\section{Varying Spectral Index} \label{appendix:varying index}
The spectra of FRBs remain poorly constrained.
There is evidence that non-repeating FRBs have wider observed bandwidths than repeating FRBs \citep[e.g.,][]{Pleunis2021}.
The CHIME sky rate at 600 MHz is $525~\mathrm{day^{-1}\,sky^{-1}}$ above $5~\mathrm{Jy\,ms}$ \citep{CHIME2021}.
After rescaling to the same fluence threshold with $N(>F_\nu) \propto F_\nu^{-1.5}$, the ASKAP rate is $\sim 440~\mathrm{day^{-1}\,sky^{-1}}$ at 1.3 GHz \citep{Shannon2018} and the Parkes rate is $\sim 430~\mathrm{day^{-1}\,sky^{-1}}$ at 1.4 GHz \citep{Bhandari2018}.
These rates do not require an extremely steep frequency dependence of the burst occurrence rate.
Ultrawideband observations of the hyperactive repeater FRB 20240114A show strong spectral variability \citep{Uttarkar2026}. Its narrowband bursts also support an interpretation in which FRB emission is not intrinsically broadband. 
The intrinsic physical bandwidth of non-repeating FRBs is still unknown.

Over the 400 MHz CHIME band, two limiting descriptions can be considered for non-repeating FRBs.
First, the emission may be intrinsically narrowband, with a distribution of central frequencies determining which observing band each burst appears in.
Second, the emission may be intrinsically broadband and approximated by a power-law spectrum.
Although physically different, these two descriptions affect the inferred redshift distribution in a similar direction.
If the event rate (for narrowband case) or rest-frame spectrum (for broadband case) declines strongly toward higher frequency, high-redshift bursts are observed at higher rest-frame frequencies and are more likely to be missed in a fixed observing band.
Correcting for this effect would shift the inferred intrinsic redshift distribution toward higher redshift.

In this Appendix, we adopt the broadband interpretation as a diagnostic test and quantify how much the inferred redshift distribution changes with spectral index.
Including the K-correction, the energy is calculated as \citep{Locatelli2019, James2022b, Shin2023}
\begin{equation}
  E = \frac{1}{(1+z)^{2+\beta}} 4\pi d_L^2 F_{\nu} \Delta \nu,
\end{equation}
where $\beta$ is the spectral index of $F_{\nu}\propto \nu^{\beta}$.
The case $\beta=-1$ reduces to the main-text energy definition (Equation~\ref{eq:energy}); in the main analysis, Equation~\ref{eq:energy} is interpreted as a band-averaged, narrowband energy estimate over the CHIME band.

The effect of varying spectral index on the inferred redshift distribution is shown in Figure~\ref{fig:Distz_baseband}.
Under the broadband assumption, previous studies have suggested a characteristic spectral index of approximately $\beta\sim -1.5$ for non-repeating FRBs \citep{Macquart2019, Shin2023}.
Changing $\beta$ shifts the inferred distribution in the expected direction, but it does not change the qualitative conclusion of this work over the range tested here.
Unless the spectral index is extremely steep, the inferred redshift distribution remains inconsistent with the SFH model.
For example, even for $\beta=-3$, the inferred distribution is still more consistent with a delayed model than with the SFH model.
Therefore, uncertainty in the spectral index does not appear to dominate the systematic uncertainty in the main conclusion for the main conclusion of this work.

\bibliography{sample701}{}
\bibliographystyle{aasjournalv7}
\end{document}